\documentclass[journal=nalefd,manuscript=article,layout=twocolumn,nano]{achemso}

\usepackage{chemformula} 
\usepackage{multirow}
\usepackage[T1]{fontenc} 
\usepackage[colorlinks = true, linkcolor = blue, urlcolor  = blue, citecolor = blue, anchorcolor = blue]{hyperref}

\usepackage{graphicx}
\usepackage{morefloats}
\usepackage{color}
\usepackage{amsmath}
\usepackage[colorlinks=true,linkcolor=blue,citecolor=blue]{hyperref}
\usepackage{soul}
\usepackage[nomarkers, nolists,figuresonly]{endfloat}

\author{Jeovani\,Brand\~ao}
\email{jeovani.brandao@lnls.br}
\affiliation{%
Laborat\'{o}rio Nacional de Luz S\'{i}ncrotron, Centro Nacional de Pesquisa em Energia e Materiais, 13083-970 Campinas SP, Brazil}%

\author{Pamela C. Carvalho}
\affiliation{Universidade de S\~ao Paulo, Instituto de F\'isica, Rua do Mat\~ao, 1371, 05508-090 S\~ao Paulo, S\~ao Paulo, Brazil}

\author{Ivan P. Miranda}
\email{ivan.depaulamiranda@lnu.se}
\affiliation{Department of Physics and Electrical Engineering, Linnaeus University, SE-39231 Kalmar, Sweden}
\affiliation{Department of Physics and Astronomy, Uppsala University, 75120 Box 516 Sweden}

\author{Thiago J. A. Mori}
\affiliation{%
Laborat\'{o}rio Nacional de Luz S\'{i}ncrotron, Centro Nacional de Pesquisa em Energia e Materiais, 13083-970 Campinas SP, Brazil}%

\author{Fanny B\'{e}ron}
\affiliation{%
Instituto de F\'{i}sica Gleb Wataghin, Universidade Estadual de Campinas,13083-859 Campinas SP, Brazil}%

\author{Anders Bergman}
\affiliation{Department of Physics and Astronomy, Uppsala University, 75120 Box 516 Sweden}

\author{Helena M. Petrilli}
 \affiliation{Universidade de S\~ao Paulo, Instituto de F\'isica, Rua do Mat\~ao, 1371, 05508-090 S\~ao Paulo, S\~ao Paulo, Brazil}

\author{Angela B. Klautau}
\affiliation{Faculdade de F\'isica, Universidade Federal do Par\'a, CEP 66075-110, Bel\'em, PA, Brazil}

\author{Julio\,C.\,Cezar}
\affiliation{%
Laborat\'{o}rio Nacional de Luz S\'{i}ncrotron, Centro Nacional de Pesquisa em Energia e Materiais, 13083-970 Campinas SP, Brazil}%

\title  {Proximity-induced flipped spin state in synthetic ferrimagnetic Pt/Co/Gd heterolayers}

\abbreviations{IR,NMR,UV}

\begin{document}

 


\begin{abstract}

To develop new devices based on synthetic ferrimagnetic heterostructures, understanding the material’s physical properties is pivotal. Here, the induced magnetic moment (IMM), magnetic exchange coupling, and spin textures are investigated in  Pt(1 nm)/Co(1.5 nm)/Gd(1 nm) multilayers using a multiscale approach. The magnitude and direction of the IMM are interpreted in the framework of both X-ray magnetic circular dichroism and density functional theory. The IMM transferred by Co across the Gd paramagnetic thickness leads to a nontrivial flipped spin state (FSS) within the Gd layers, in which their magnetic moments couple antiparallel/parallel with the ferromagnetic Co near/far from the Co/Gd interface, respectively. The FSS depends on the magnetic field, which, on average, reduces the Gd magnetic moment as the field increases. For the Pt, in both Pt/Co and Gd/Pt interfaces, the IMM follows the same direction as the Co magnetic moment, with negligible IMM in the Gd/Pt interface. Additionally, zero-field spin spirals were imaged using scanning transmission X-ray microscopy, whereas micromagnetic simulations were employed to unfold the interactions, stabilizing the ferrimagnetic configurations, where the existence of a sizable Dzyaloshinskii-Moriya interaction is demonstrated to be crucial.  

\end{abstract}

\vspace{5mm}

\textbf{Introduction}

\vspace{5mm}
Interface-induced properties in thin films and multilayers have been considered to play an important role in different processes,  including spin transport, interfacial anisotropy, topological superconductivity, and proximity effect \cite{PhysRevLett.109.107204, Hellman2017, Nguyen, Zhang}. The latter is known to transfer electronic properties to other materials where it is not present. For example, the surface/interface of a certain material without electronic order may acquire superconductivity and ferromagnetism in contact with superconductors or ferromagnets, respectively \cite{RevModPhys.77.935,PhysRevLett.105.077001, PhysRevB.93.155402, PhysRevB.55.3663, PhysRevLett.85.413}. So far, particularly, induced magnetic moment (IMM) has been investigated mostly in ferromagnetic/non-magnetic interfaces, where the magnetization of $3d$ transition metals (Co, Fe, Ni) induces a magnetic polarization in $4d$ (Pd), $5d$ heavy metals (W, Ir, Pt) when they are thin and in proximity \cite{PhysRevB.60.12933,PhysRevLett.87.207202, PhysRevB.90.104403}.

The relationship between IMM and different interface/transport phenomena has been explored in several systems. To cite some examples, IMM and its role have been investigated in topological insulators from low to room temperature, allowing the use of magnetic doped insulators for spintronic devices \cite{doi:10.1021/nl201275q}. IMM and interfacial Dzyaloshinskii-Moriya interaction (DMI) has also been explored in Pt/Co systems, revealing that by inserting Au or Ir space layers between the Co/Pt interface, IMM decreases rapidly and vanish, while DMI remains \cite{Rowan}. Spin pumping and IMM have also been explored in ferromagnetic/nonmagnetic interfaces, and their impact on scattering of pure spin current in electrical transport remains under debate \cite{PhysRevLett.109.107204,PhysRevB.94.014414, PhysRevB.98.134406, PhysRevB.99.064406}.

Only in the last few years IMM has gathered more attention in ferrimagnetic systems \cite{ PhysRevLett.110.147207, 10.1063/1.4987145, PhysRevResearch.2.033280}  due to emergent phenomena, as exemplified by all-optical switching (AOS) \cite{PhysRevLett.99.047601, PhysRevB.96.220411}, suppression of the skyrmion Hall effect \cite{natnanotech, natcomm}, faster domain wall motion \cite{PhysRevLett.121.057701}  and THz dynamics \cite{PhysRevLett.108.247207, PhysRevLett.127.037203}. Pulsed laser driving magnetization reversal, and topological spin textures nucleation such as skyrmions, has been observed in ferrimagnetic films even without the assistance of magnetic fields \cite{AOS-Nature,AOS-Natcomm, doi:10.1021/acsami.2c19411,D3NR04529C, doi:10.1021/acsaelm.3c01534}. These mechanisms place ferrimagnets as an energy-efficient material for nonvolatile ultrafast toggle switching to develop antiferromagnetic spintronics devices \cite{Refe01, Refe02, Refe03}.

Thin films and multilayers composed of Co and Gd alloy or bilayers made of Co/Gd are examples of ferrimagnetic coupling where the Co $3d$  transition metal and Gd $4f$ rare-earth magnetic moments may align antiferromagnetically regarding each other depending on the composition and temperature \cite{PhysRevB.74.134404}.  This antiparallel alignment emerges due to an indirect negative exchange played by the bridge role, which essentially means that the Gd $5d$ and $4f$ magnetic moments are parallel via direct exchange (or via a polarizing field generated by the localized $f$ states, when they are theoretically treated as part of the core), however, are antiparallel to the Co $3d$ magnetic moments through the hybridization of the itinerant $5d-3d$ states. \cite{Lee}.

When Pt is added as an under/over layer in  CoGd alloys or Co/Gd bi-layer, its IMM physical understanding remains under debate. Pt is a heavy paramagnetic material with an electronic band structure that may satisfy the Stoner criterion to acquire ferromagnetic order. Therefore, Pt does not exhibit spontaneous magnetic polarization, except when occurs size reduction or proximity with a ferromagnetic material \cite{Pt.Ferromag, FLORIAN}. Besides, an important question arises about the magnetic polarization for Pt in contact with rare-earth elements such as Gd, which calls attention to the role played by Gd-Pt ($5d-5d$ and $4f-5d$) interaction in transferring  IMM. For Pt/CoFeGd interfaces, where the CoFeGd layer is deposited as alloy, the IMM  transferred for Pt comes from the 3d magnetic metal even below or above the compensation temperature \cite{PhysRevResearch.2.033280}. On the other hand, it has been argued experimentally and theoretically that both Gd and CoFe subllatices work in phase to transfer their respective magnetic moments to the Pt \cite{SRP-proximity}.  To quantify the IMM and its implication on the interfacial magnetic properties of various phenomena occurring in multilayered materials, advanced experimental techniques with chemical selectivity in combination with theoretical models are key tools to identify the direction, magnitude, and effects of the IMM.

In this work, the element-resolved magnetic moment was probed in Pt/Co/Gd heterolayers. Using XMCD, magnetic moments were detected individually for each element. IMM transferred from Co to Pt and Gd was measured at room temperature. From the average XMCD signal, experimentally, both magnetic moments in Co and Pt are aligned parallel with the external magnetic field direction. However, the magnetic polarization in the Gd is antiparallel with respect to the magnetic field. Using well-known sum rules the magnitudes of the orbital and spin magnetic moments were experimentally determined, and to compare with theoretical values, density functional theory was employed. The theoretical results resolved layer-by-layer for each element show that the IMM in the Gd layer near the Co/Gd interface is coupled antiferromagnetically with the Co, whereas the Gd layers far from the Co/Gd interface are coupled ferromagnetically.  Thus, herein this behavior designated as a flipped spin state (FSS) stands as an interface effect where an antiferromagnetic coupling occurs between interfacial Co and Gd atoms. The FSS also depends on the magnetic field amplitude and occurs at all investigated temperatures.

This contrasts with the ferromagnetic configuration observed within the Gd below the Curie temperature ($T_{\textnormal{C}}$) in the absence of Co, in which FSS does not appear. This theoretical achievement of FSS within the Gd layer is further reinforced by comparing the average experimental and theoretical magnetic moments. This finding calls attention to the need to understand both IMM and FSS in interface-related phenomena and electric transport in multilayered thin films.  In addition, scanning transmission X-ray microscopy reveals the formation of antiferromagnetic coupled spin spirals as the magnetic configurations at the nanoscale in the multilayer. Micromagnetic simulations were performed to reproduce the experimental images and understand the energy landscape leading to such spin textures.

\vspace{5mm}

\textbf{Results and Discussion}

\textbf{Magnetic properties of base Pt/Co and Pt/Gd multilayers.} To investigate the IMM transferred from Co to both room-temperature paramagnetic Pt and Gd, two reference multilayers based on Pt/Co and Pt/Gd were grown by magnetron sputtering deposition onto Si/SiO$_{2}$ substrate, see Methods and Supplementary Note 1. By means of hysteresis loops, as expected, only the Pt/Co multilayer shows ferromagnetic behavior, whereas the Pt/Gd one presents a paramagnetic signature. Further XAS and XMCD measurements also confirm the different magnetic phases; see Supplementary Figure 1.

It is worth anticipating that simulations for the Pt/Co/Pt multilayers were performed to establish the methodology used for comparing theoretical and experimental results, since this system is well-known in the literature. Through first-principles calculations and atomistic spin dynamics (ASD), good agreement was obtained for the atomic Co magnetic moment (for more details, see the Supplementary Note 3).

\vspace{5mm}

\vspace{5mm}

\begin{figure*}[h!]
\centering
 \includegraphics[scale=0.5]{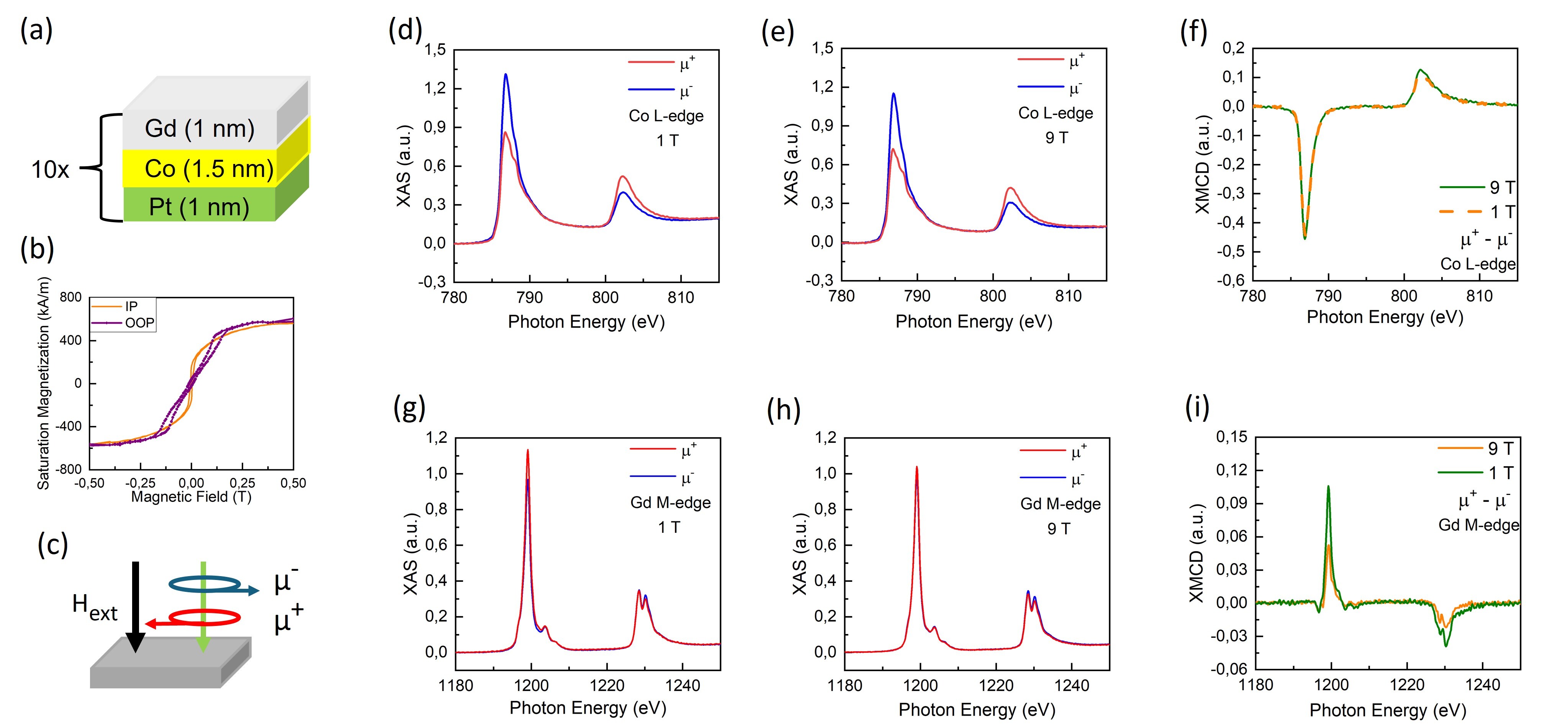}
\centering
\caption{\textbf{Sample stack, hysteresis loops, and XAS/XMCD geometry and measurements.} In (a) schematic representation of the Pt/Co/Gd heterolayer, corresponding to one stacking unit (to be repeated $10\times$). In (b)  hysteresis loops  acquired for both in-plane (IP) and out-of-plane (OOP) magnetic fields. X-ray absorption (XAS) and X-ray magnetic circular dichroism XMCD geometries applied in the measurements  displayed in (c), where $\mu^+$ ($\mu^-$) denote the right and left circularly polarized X-rays, respectively, and $H_{\textnormal{ext}}$ the external applied magnetic field.  In (d) and (e)  XAS acquired for right and left circularly polarized X-rays around the Co absorption edges for 1 and 9 T magnetic fields. In (f), the resultant XMCD for both magnetic fields. XAS  measured  around the Gd absorption edges for 1 and 9 T magnetic fields  are shown in (g) and (h), respectively. The resultant XMCD displayed in (i) shows different amplitudes in comparison for each field. }
\label{fig2}
\end{figure*}

\textbf{Induced magnetic moment in rare-earth Gd layers.} The IMM was investigated first at room temperature in rare-earth Gd by introducing a thin Co (1.5 nm) thickness, thus forming a Pt/Co/Gd multilayer. 
Figure~\ref{fig2}(a) shows a schematic representation of the Pt(1nm)/Co(1.5nm)/Gd(1nm) stack. The three layers were repeated 10 times; thus, above Gd there is Pt forming Pt/Co and Gd/Pt interfaces in the heterostructure. X-ray diffraction was measured to verify the crystalline quality (Supplementary Figure 2(a)), indicating that the sample formed a textured heterostructure with a preferential growth direction along the [111] axis of the metallic layers. X-ray reflectivity (XRR) measurements and simulations (Supplementary Figure 2(b-d)) were performed to confirm the multilayer stacking morphology. The XRR best-fit model indicates that the sequential deposition of the Pt/Co/Gd trilayer indeed forms a heterostructure rather than just a Pt-Co-Gd alloy. Although the XRR analysis cannot distinguish between interdiffusion and roughness at the interfaces, the scattering length density (SLD) profile demonstrates that each layer has regions without any mixing effects. An estimate of the mixing thickness around the interfaces is highlighted in Supplementary Figure 1(d).  

\begin{figure*}[!ht]
\centering
 \includegraphics[scale=0.5]{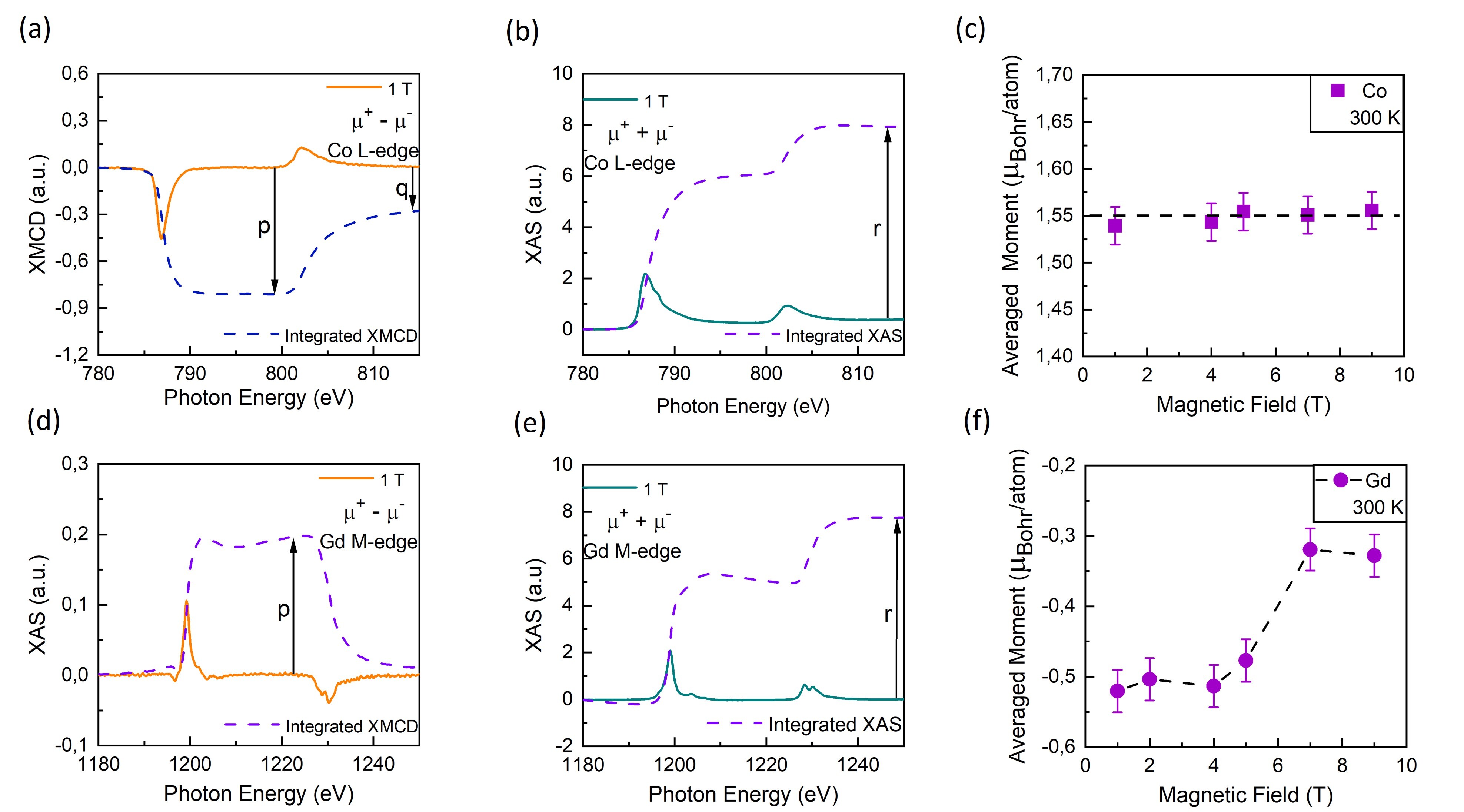}
\centering
\caption{\textbf{Integrated XAS and XMCD for both Co and Gd elements applied in the sum rules.} In (a) and (d) are shown the X-ray magnetic circular dichroism (XMCD) for both Co and Gd, and the calculated $p$ and $q$ parameters. The parameter $q$ is not represented in (d) due to its small value. X-ray absorption spectra (XAS) acquired for right  ($\mu^{+})$ and left   ($\mu^{-}$) circularly polarized X-ray are summed ($\mu^{+}$ + $\mu^{-}$) for Co and Gd and displayed in (b) and (e), correspondingly. The integrals taken over the spectra are used to estimate the amount $r$. The discussion on the extracted parameters is detailed in the Supplementary Note 1. Quantities $p$, $q$, and $r$ are used to extract orbital and spin magnetic moments in the sum rules equation and hence the averaged (orbital + spin)  moments. The averaged  moments are strictly constant as a function of the magnetic field, as shown in (c). For the Gd, in module, the averaged  moments are constant from 1 to 5 T, and then reduces drastically for fields between 8 and 9 T as observed in (f). The error bars were estimated by the values of the parameters $p$, $q$ and $r$ extracted along with the XMCD and XAS.} 
\label{fig3}
\end{figure*}

Figure~\ref{fig2}(b) shows the hysteresis loops measured with in-plane and out-of-plane magnetic fields. From the reversal magnetization driven by the magnetic field, it is possible to infer that the heterostructure exhibits perpendicular magnetic anisotropy (PMA) because of the smaller magnetic field to saturate the sample magnetization normal to the surface. The in-plane remnant magnetization observed in the hysteresis loop could be due to spin reorientation, which occurs in similar systems when the Co thickness increases \cite{PhysRevB.97.174419}.

The individual magnetism of Co and Gd was probed by XAS and XMCD in the Soft X-ray Absorption and Imaging (SABIÁ) beamline of the new 4th synchrotron generation Sirius. The magnetic fields were applied perpendicular to the sample surface and parallel to the incoming X-ray by using a superconducting magnet, see in Figure~\ref{fig2}(c) the schematic geometry used in the XAS/XMCD measurements. The signal was recorded in the total electron yield (TEY) mode, collecting the  drain current from ground to sample, (see more detail in Methods). Figure ~\ref{fig2}(d) and (e) depict clearly the dependent XAS concerning the X-ray polarization ($\mu^{+}$ and $\mu^{-}$) examined around the L$_{2,3}$ Co absorption edges, for both 1 and 9 T magnetic fields. In Figure ~\ref{fig2} (f), the respective XMCDs for 1 and 9 T were plotted together. Each XMCD points down/up around the L$_{3}$ and L$_{2}$ edges, and exhibit a similar magnitude.

The same measurements were carried out for the Gd layer. Figure~\ref{fig2}(g) and (h) show the spectra acquired around the M$_{4,5}$ Gd edges. For 1 and 9 T, the XAS exhibit  difference between the spectra taken with right and left circular polarization ($\mu^{+}$ and $\mu^{-}$), which is clearly observed in the averaged XMCD shown in Figure~\ref{fig2}(i).  This difference shed lights on the average magnetic moment within the Gd layer, which depends on the applied magnetic field. Moreover, contrary to what was observed in the Co layer, the average XMCD points downward/ upward around the M$_{5,4}$ Gd edges. This means that the average Co and Gd magnetic moments are parallel/antiparallel to the magnetic field. 


\textbf{Co and Gd atomic magnetic moments evaluated by sum rules.} To assess and separate the orbital and spin magnetic moments for both Co and Gd layers, the well-known sum rules were employed by integrating the XAS and XMCD spectra for each element -- see details in the Supplementary Note 1.

Figures~\ref{fig3}(a,d) and (b,e) show the XMCD and XAS along with  its respective integrated signals for Co and Gd. Specifically, for the integral performed over the Gd XMCD spectra, the parameter $q$ could not be represented since the integral goes almost to zero (Figure~\ref{fig3}(d)), which may be due to the half full Gd $4f$ shell \cite{orbital}. The others parameters $p$ and $r$ were also used in the sum rules; see Supplementary Note 2. For the Co, and both applied 1 and 9 T magnetic fields,  the average orbital and spin magnetic moments were quantified as $0.14 \pm 0.05$ and   $1.41\pm 0.05$ $\mu_{B}$$\cdot\textnormal{atom}^{-1}$. On the other hand, for the Gd, in the applied  1 T magnetic field the orbital and spin magnetic moments were  $-0.04\pm 0.01$  and $-0.54 \pm 0.05$ $\mu_{B}$$\cdot\textnormal{atom}^{-1}$, while   for 9 T,   $-0.015\pm 0.005$  and $-0.31 \pm 0.05$ $\mu_{B}$$\cdot\textnormal{atom}^{-1}$. Consequently, the Gd magnetic moment is, on average, reduced for higher magnetic fields.

\begin{figure*}[!ht]
\includegraphics[scale=0.6]{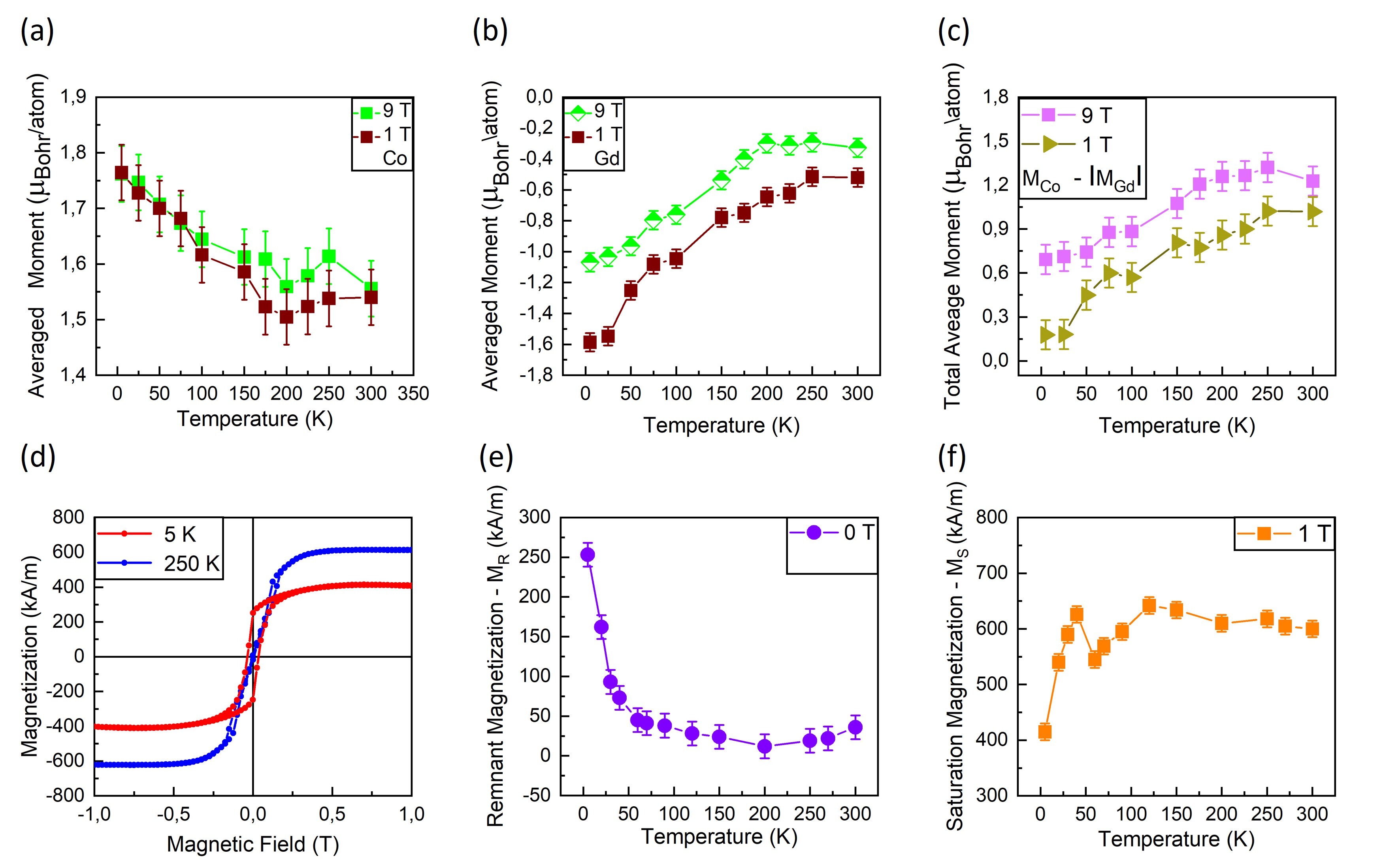}
\caption{\textbf{Temperature-dependent magnetic moments by XMCD and SQUID.} In (a),  Co magnetic moments reduce as the multilayer is warmed-up from 5 to 300 K. In (b),  Gd presents similar behavior where, (in magnitude), the moments are reduced. Note that there is in average for the Gd a difference  between  the amplitudes of the magnetic moments in comparison with the 1 and 9 T fields. In (c), the total magnetic moments  acquired by using (M$_{Co}$ - |M$_{Gd}$|) . The total magnetic moments for both 1 and 9 T decrease as the temperature reduces. The lines are guide to eyes and the error bars obtained from the  parameters $p$, $q$ and $r$ extracted along with the XMCD and XAS.  Hysteresis loops measured with field applied in the  out-of-plane at 5 and 250 K, (d). In (e) the remnant magnetization extracted from the hysteresis loops in the absence of magnetic field in function of the temperature. (f) shows the saturation magnetization normalized by the sample volume for different temperatures under applied magnetic field of 1 T. {The error bars were obtained from the remnant and saturation magnetization extracted for both positive and negative magnetic fields. }}
\label{fig3b}
\end{figure*}

In a previous work performed for a comparable  Pt/Co/Gd multilayer, where the Co thickness was 0.5 nm, magnetic moments were determined by means of XAS and XMCD spectra acquired at zero-field. Even so, the magnitudes of the magnetic moments  for Co and Gd were found to be of the similar order as those obtained in our work, with differences attributable to the remnant state of the sample at different temperatures and absence of applied magnetic fields \cite{PhysRevMaterials.6.084412}.

Despite the magnetic moments values, it calls attention that for the  Co, the magnetic moment does not depend on the magnetic field, while for the  Gd as the magnetic field increases, it is weakened. This field-dependent magnetic moment for the Co is shown in Figure~\ref{fig3}(c), and is constant within the experimental error. For the Gd, the average moment (in magnitude) is higher from 1 up to 5 T and almost constant considering the experimental error. Suddenly, the average moment decreases for fields greater than 5 T, Figure~\ref{fig3}(f). It appears that there is a critical field H$_{C}$ between 5 and 7 T, at which the internal magnetization in the Gd layer changes in such a way that the average magnetic moment drops. Enlarging the magnetic field further to 9 T, the average magnetic moments are constant.

This implies that the IMM within the Gd layers is about one-third of the Co magnitude when the field is 1 T, and almost one-fifth for 9 T. This finding highlights the range in which the magnetic moments of Co can induce polarization across the Gd thickness, and the nontrivial arrangement of the magnetic moments within the Gd layer. This aspect will be later inspected through \textit{ab-initio} calculations.  

\begin{figure*}
\centering
 \includegraphics[scale=0.6]{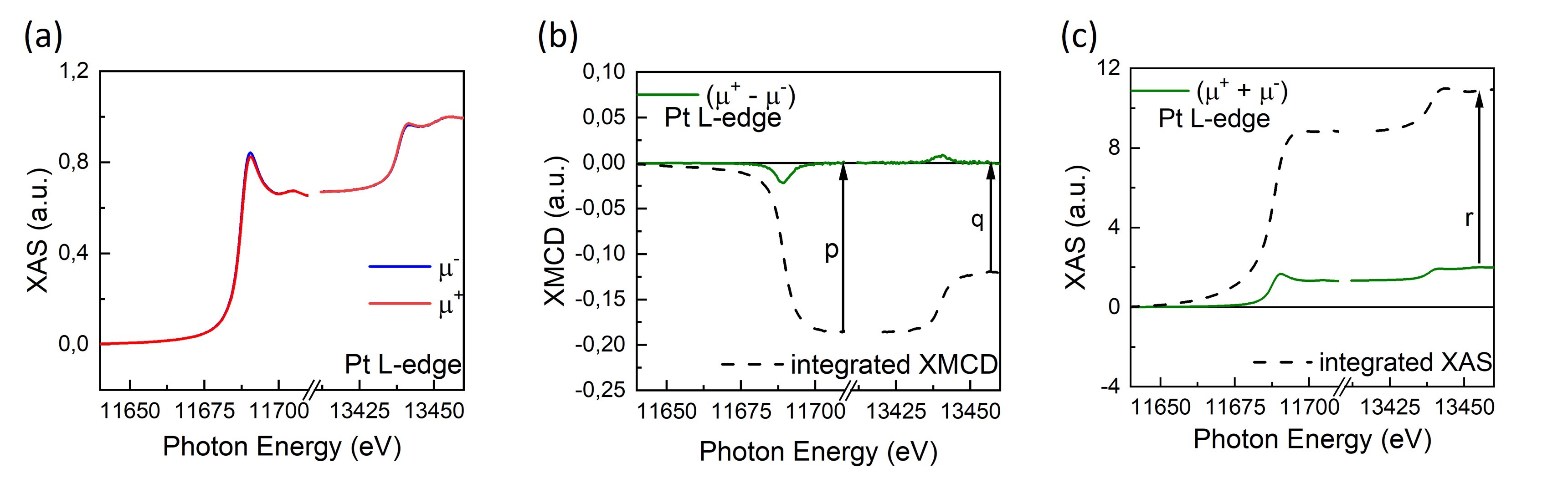}
\centering
\caption{\textbf{XAS and XMCD performed around the Pt L$_{2,3}$ edges.} In (a),  X-ray absorption spectra (XAS) acquired for right ($\mu^+$, red solid line) and left ($\mu^-$, blue solid line) circularly polarized X-rays show the difference between the spectra. In (b) the resultant X-ray magnetic circular dichroism (XMCD), given by $\mu^{+} - \mu^{-}$ (solid green line), and the integrated signal (dashed line) over the spectra. In (c) the quantity $\mu^{+} + \mu^{-}$ and its respective integrated signal. The parameters $p$, $q$ and $r$ are used to perform sum rules and extract the magnetic moment.}
\label{fig4} 
\end{figure*}

\vspace{5mm}

\textbf{Co and Gd temperature-dependent  magnetic moments.} Co and Gd magnetic moments were quantified  across the temperature by XAS/XMCD.  The  spectra were measured from 5 to 300 K, and for both 1 and 9T magnetic fields. Figure~\ref{fig3b} resumes the temperature-dependent magnetic moments. In Figure~\ref{fig3b}(a), the Co magnetic moment decays  from 5  to 200 K, and then increases slightly from 200 to 300 K. For both 1 and 9 T magnetic fields, the magnitude of the magnetic moments are very similar and do not show substantial difference covering  the whole temperature range.

In Figure~\ref{fig3b}(b), the Gd magnetic moment decreases (in magnitude) as the temperature increases from 5 to 300 K. Contrary to the observed for the Co, there is a difference between the magnetic moments quantified for 1 and 9 T where the moment, on average, is smaller for 9 T.  It also seems that the amplitude of the Gd magnetic moments reduces progressively  from 5 to 300 K independent of the applied magnetic field.

The resultant magnetic moment as a function of the temperature  was obtained by subtracting (M$_{Co}$ - |M$_{Gd}$|), see Figure~\ref{fig3b}(c). The total magnetic moment  increases from 5 to 300 K for both fields, with magnitude smaller for 1 T. This can be explained by the higher average magnetic moment hosted for the Gd at 1 T in all temperatures in comparison with 9 T, and hence  the complex internally flipped spin state surrounded by the Gd layer.

SQUID measurements were also performed to account for the magnetization of the entire multilayer. Hysteresis loops were acquired with the magnetic field of 1 T applied out-of-plane. In Figure~\ref{fig3b}(d), is shown examples obtained at 5 and 250 K. It is clearly observed that the hysteresis loops have different shapes. From the hysteresis loops were extracted the remnant and saturation magnetization in function of the temperature.

In Figure~\ref{fig3b}(e), the remnant  magnetization reduces rapidly from 5 to around 75 K, then continues slowly reducing up to 200 K. From this temperature there is a slight increase up to 300 K. It appears that around 200 K the magnetization reaches its lowest value, although it is not possible to define an exact compensation temperature T$_{M}$. An important aspect on the remnant magnetization lies in the thermally stable temperature region ($\approx$ 200 to 300 K), where this property does not drastically change.

The saturation magnetization as a function of the temperature obtained for 1 T is presented in Figure~\ref{fig3b}(e). From low to high  temperatures the saturation magnetization increases, and reaches an almost stable value above 150 K. The trend is similar to what was observed for the magnetic moment extracted from the sum rules shown in Figure~\ref{fig3b}(c).

\textbf{Induced magnetic moment at the Pt/Co and Gd/Pt interfaces.} To fully understand IMM in multilayer, XAS and XMCD were also performed around the Pt L$_{2,3}$ absorption edges in the Pt(1nm)/Co(1.5nm)/Gd(1nm) multilayer. XAS was measured in the hard X-ray energy range at the Extreme condition Methods of Analysis (EMA) beamline of the Sirius. Right and left circularly polarized X-rays were generated by a 1/4 wave plate, and a magnetic field of 1 T perpendicular to the sample was applied during the  XAS measurements. The signal was recorded in the fluorescence mode. 
 
Figure~\ref{fig4}(a) shows the XAS acquired for both incoming right and left circular polarization.  It shows a difference in absorption around both L$_{2}$ and L$_{3}$  edges, which is observed in the XMCD signal in Figure~\ref{fig4}(b). The XMCD points downward/upward around the  L$_{3}$ and L$_{2}$ absorption edges, respectively. Figure~\ref{fig4}(c) exhibits the sum of the XAS for right- and left-circularly polarized X-ray. Sum rules were also used to quantify the magnetic moment in the Pt layer. The average orbital and spin magnetic moments were obtained as 0.026 $\pm$ 0.01 $\mu_{B}$$\cdot\textnormal{atom}^{-1}$ and 0.09  $\pm$ 0.04 $\mu_{B}$$\cdot\textnormal{atom}^{-1}$, respectively. Our results are comparable with previous work reporting IMM in Pt layers at  Pt/Co, Pt/Ni and Pt/Y$_{3}$Fe$_{5}$O$_{12}$ interfaces  \cite{Pt1,Pt2,Pt3}. Indeed, as will be demonstrated, our theoretical results based on atomistic simulations confirm that the IMM transferred for Pt comes predominantly from the Co layer rather than the Gd.

\vspace{5mm}

\textbf{Theoretical analysis: Atomistic modeling.} To uncover the physical aspects driving the IMM into the Gd and Pt layers at the interface, the Pt/Co/Gd system was investigated by a combination of first-principles calculations based on density functional theory (DFT) and atomistic spin dynamics (ASD) simulations for different temperatures (see Methods section).

\begin{figure*}
\centering
 \includegraphics[scale=1]{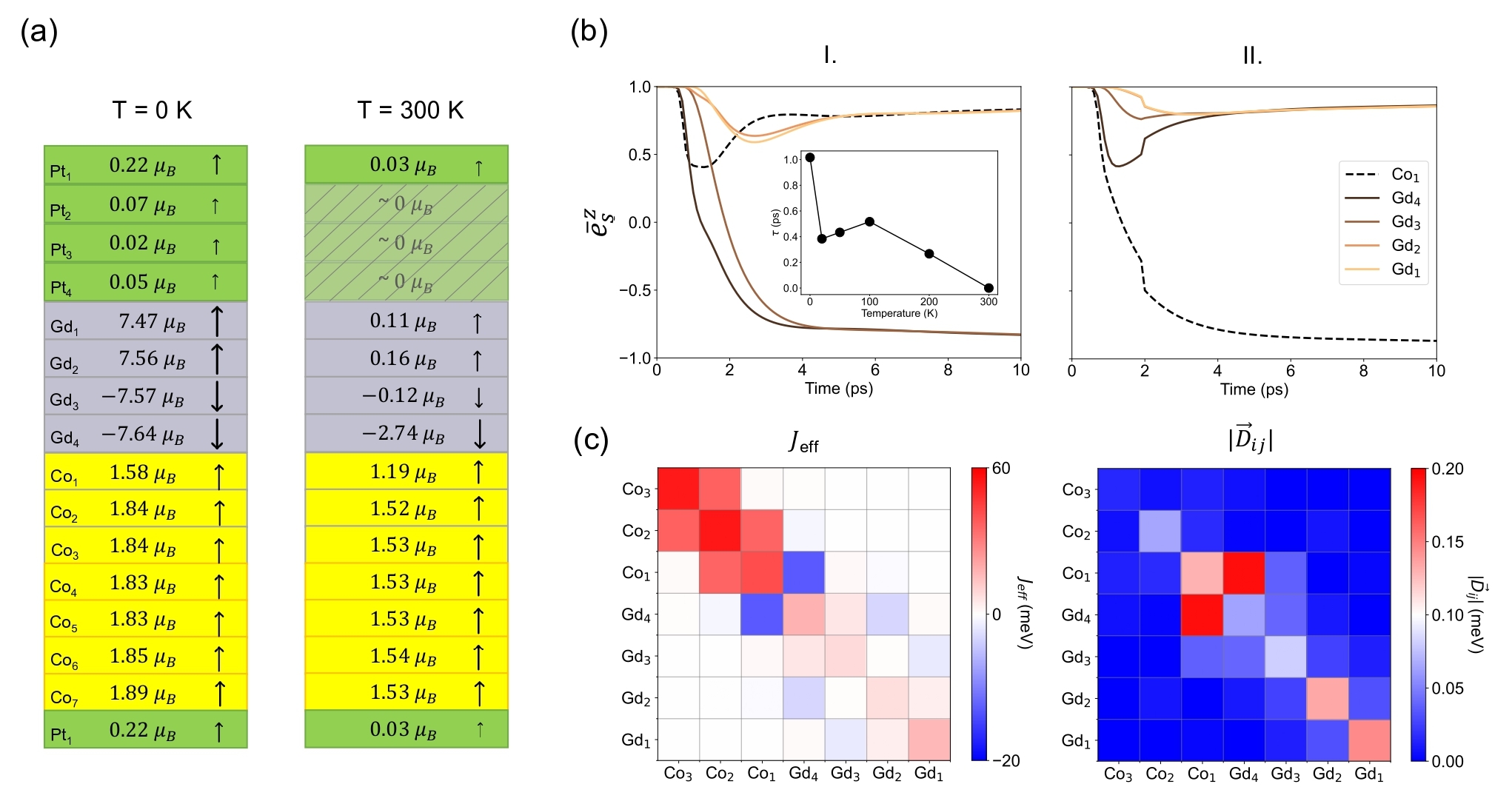}
\centering
\caption{\textbf{Density functional theory and atomistic spin dynamics results for Pt/Co/Gd.} In (a), the pristine Pt/Co/Gd stacking is illustrated where the green, gray, and yellow layers represent the Pt, Gd, and Co atoms, respectively. The atomic spin magnetic moments for each layer are indicated along with the direction of polarization (arrow) for 0 K and 300 K. The first Pt layer representation (top) is repeated at the bottom to show the connection with the upcoming Co$_7$ layer. In (b), the $\hat{z}$-component of the spin moment direction ($\bar{e}^{z}_s$) is plotted as a function of time for the Co$_{1}$, Gd$_{4}$, Gd$_{3}$, Gd$_{2}$ and Gd$_{1}$ layers. Here, the case with the full set of magnetic couplings (Co$_{1}$-Co$_{1}$, Co$_{1}$-Co$_{2}$, Gd$_{4}$-Co$_{1}$, etc) is considered in I., while in II. the Gd$_4$-Gd$_1$ and Gd$_4$-Gd$_2$ interlayer couplings are set to zero. In (c), the effective exchange interactions ($J_{\textnormal{eff}} = \sum_{i \neq j}J_{ij}$) and strength of the nearest-neighbor Dzyaloshinskii-Moriya interaction ($|\vec{D}_{ij}|$) are represented in a color map for different couplings among Gd and the nearest Co layers. The color bars in (c) represent the magnitude of $J_{\textnormal{eff}}$ and $|\vec{D}_{ij}|$ in meV, as obtained via first-principles calculations, from minimum (in blue) to maximum (in red).
}
\label{moms-dft}
\end{figure*}

The layer-resolved atomic spin magnetic moments ($\mu_{s}$), (for the pristine structure which considers the system without defects), are displayed in Figure~\ref{moms-dft}(a), and the corresponding static average spin ($\bar{\mu}_{s}^z$) magnetic moments for the Pt, Co, and Gd atoms for $T=0$ K are presented in Table~\ref{tab:moms-cogd}. At the ground-state, our results show that the Gd dominates the magnetism in absolute values, presenting a spin magnetic moment very close to its measured bulk value (7.63 $\mu_{B}$$\cdot\textnormal{atom}^{-1}$ \cite{JensenMackintosh1991}).
As can be seen in Figure~\ref{moms-dft}(a), within the Gd thickness, the two layers closest to the Co/Gd interface align antiferromagnetically to the Co spin state, while the two layers farthest to that interface present a ferromagnetic alignment with Co, characterizing the FSS behavior. 
This reduces the overall ground-state average magnetic moment of Gd to almost zero (see Table~\ref{tab:moms-cogd}) for the pristine structure. Indeed, the temperature-dependent magnetic moment extracted for Co and Gd summarizes this behavior, in which the resultant magnetic moment for the Gd at $T\rightarrow0$ tends to vanish.
It is, thus, an effect of the interface with Co, since, alone, Gd tends to show a purely ferromagnetic behavior below its Curie temperature and high magnetic moment about 7 $\mu_{B}$$\cdot\textnormal{atom}^{-1}$ at low temperatures \cite{Gd-moment}.

 \begin{table*}[h]
 \centering
  \caption*{Theoretical and experimental average spin magnetic moments.}
    
   \begin{tabular}{c|cc|c|cc|cc}
    \hline 
         & \multicolumn{3}{c|}{\textbf{Theoretical}} & \multicolumn{2}{c|}{\textbf{Experimental (1 T)}} & \multicolumn{2}{c}{\textbf{Experimental (9 T)}}  \\ \hline
        &  \multicolumn{2}{c|}{$T = 0$ K}  &  $T = 300$ K & $T = 5$ K & $T = 300$ K & $T = 5$ K & $T = 300$ K   \\ 
         & P & I-R &  P  &   &   &   &     \\ 
        \textbf{Pt} & 0.09 &0.15 & 0.03  & ---& 0.09 & --- & --- \\
        \textbf{Co} & 1.85 & 1.88 & 1.48  & 1.57 &   1.41 & 1.57 &  1.41  \\
        \textbf{Gd} & $-0.18$\textsuperscript{\emph{a}} & $-0.49$ & $-0.67$  & $-1.48$  & $-0.51$ & $-0.95$ & $-0.31$ \\ \hline \\
    \end{tabular}
    \caption{\small{Theoretical ($T = 0$ and 300 K) and experimental ($T =  5$ and 300 K) average spin ($\bar{\mu}_{s}$) magnetic moments for Pt, Co and Gd atoms in the Pt/Co/Gd multilayer, in $\mu_{B}$$\cdot\textnormal{atom}^{-1}$. Negative/positive signs indicate distinct averaged moment directions projected in the $\hat{z}$-axis. The minus sign denotes an antiparallel spin orientation. (P) denotes theoretical results considering no imperfections at the interfaces (pristine) and (I-R) for calculations considering both interdiffusion and roughness at interfaces (see Supplementary Note 3 for details). Here, the uncertainty for the experimental results is 0.05 $\mu_{B}$$\cdot\textnormal{atom}^{-1}$.}}
    \emph{a} Note that $\bar{\mu}_{s}\rightarrow0$ due to the almost antiferromagnetic spin configuration throughout the Gd thickness. However, absolute values amount to $\sim7.6\,\mu_{B}$$\cdot\textnormal{atom}^{-1}$ (see Figure \ref{moms-dft}(a)).
    \label{tab:moms-cogd}
\end{table*}

While the Co spin magnetic moment slightly increases at the Co/Pt interface, it decreases at the interfacial region close to Gd. It is worth noting that the Co atoms are responsible for the IMM in the Pt layers, with a significant induced moment of $\mu_{s}=0.22\mu_{B}$$\cdot\textnormal{atom}^{-1}$ at the Co/Pt interface, while the IMM at the Pt/Gd interface is negligible ($0.05\mu_{B}$$\cdot\textnormal{atom}^{-1}$). This can be explained by the fact that the hybridization between Pt and Gd atoms occurs prominently between Pt $5d$ and Gd $5d$, $6p$ and $6s$ valence states, while the Gd 
$4f$ states, here taken into account by an open-core treatment ($7\mu_{B}$$\cdot\textnormal{atom}^{-1}$), are highly localized. Because of this fact, even when the $4f$ states are explicitly taken into account, and treated with a Hubbard $U$ approach~\cite{Dudarev1998}, the induced Pt spin moment by Gd is still very small (as discussed in the Supplementary Note 3). Therefore, based on first-principles calculations, it allows to infer that the IMM in the Pt layers follows the Co magnetic moment, presenting a low influence of the Gd atoms. It is also possible to anticipate that this scenario might not be significantly changed by temperature effects.

In our study, the closest comparison between the theoretical achievements at $T = 0$ K (ground state)  and the experimental results can be done for the measurements acquired at $T =  5$ K, which was the lowest temperature reached in the superconducting magnet, see  Table~\ref{tab:moms-cogd}. For both 1 and 9 T magnetic fields, despite the minimal difference in temperature (0 and 5 K), the Co average spin magnetic moments were experimentally evaluated as 1.57 $\mu_{B}$$\cdot\textnormal{atom}^{-1}$, which in comparison with the theoretical result 1.85 $\mu_{B}$$\cdot\textnormal{atom}^{-1}$ (pristine case), is $\sim$ 16\% smaller. In part, such slight discrepancy may be attributed to the quench of the Co spin moment when it is intermixed in the Gd layer (see Supplementary Note 3). Nevertheless, the results capture with reasonable  magnitude, both theoretical and experimental Co spin magnetic moment at low temperature.

Turning now to Gd, the theoretical average spin magnetic moment at $T = 0$ K is $-0.18$ $\mu_{B}$$\cdot\textnormal{atom}^{-1}$ (pristine case), while experimentally at $T = 5$ K for 1 and 9 T magnetic fields, quantified as $-1.48$ and $-0.95$ $\mu_{B}$$\cdot\textnormal{atom}^{-1}$, respectively. First, we notice that the experimental values are far from reaching the expected spin moment of around $\sim7$ $\mu_{B}$$\cdot\textnormal{atom}^{-1}$, commonly predicted for Gd at low temperatures \cite{Gd-moment}, which would indicate a trivial state, i.e., ferromagnetically ordered within the Gd layers. Thus, the discrepancy may indicate a more complex (non-trivial) magnetic moment distribution within the Gd layers, reinforcing the existence of a single (or even multiple) flipped spin state(s) across the rare-earth layer.

Several hypotheses can be formulated to explain the difference between the theoretical and experimental values of the Gd spin magnetic moments: (\textit{i}) the ordered Gd magnetic moments at or far from the Co/Gd interface may be canted due to the exchange interaction between the atomic Gd layers and to the applied magnetic field (usually referred to as spin-flop transition \cite{Chen2022,Becker2017}), forming a noncollinear FSS; 
(\textit{ii}) the presence of intermixing regions may allow for the coexistence of multiple flipped magnetic states (see Supplementary Note 3), which can become close in energy; (\textit{iii}) long-range Ruderman-Kittel-Kasuya-Yosida (RKKY) interactions may lead to an oscillatory magnetic behavior within the non-intermixed Gd layers, breaking the balance of the collinear Gd spin polarizations in the ideal FSS state, resulting in a non-null average Gd spin magnetic moment; (\textit{iv}) the formation of a partial CoGd alloy near the interface (see discussion in Supplementary Note 3); and (\textit{v}) an imbalance on the amount of intermixed Gd atoms, ferro- or antiferromagnetically coupled at the interfaces (and throughout the Gd thickness). 

As such, the possibility of interfacial alloying and/or roughness was  theoretically investigated at the Pt/Co/Gd stacking at $T = 0$ K (see complete discussion in Supplementary Note 3). For example, if one considers the same number of Gd atoms interdiffused on Co and Pt layers at the FSS state, the magnetization per atom of the system is not substantially affected by the presence of atomic intermixing and/or vacancies (see Supplementary Table 1 and Supplementary Figure 8, cases A and B), where, compared to the pristine system (Figure~\ref{moms-dft}(a)), the average spin magnetic moment varies only $\sim 1.5\%$ and $\sim2\%$ for the Co and Gd atoms, respectively. Conversely, in a simplified scenario (case (\textit{v})) where there is an extra Gd atom at the Co-Gd interface compared to the Gd-Pt interface (as detailed in Table \ref{tab:moms-cogd}, Supplementary Table 1, and Supplementary Figure 8, case C), the theoretical average Gd spin magnetic moment is calculated to be $-0.49$ $\mu_{B}$$\cdot\textnormal{atom}^{-1}$ at $T = 0$ K. Thus, this provides one possible explanation for the residual Gd magnetic moment observed experimentally at low temperatures.


Nevertheless, if the amount of Gd atoms interdiffused in Co layers is greater compared to the Gd-Pt interface, this imbalance in the number of Gd atoms ferro- and antiferro-magnetically coupled at the interfaces (and throughout the Gd thickness) results in a residual Gd AFM moment coupled to Co. The simplest scenario with this characteristic, where there is an additional Gd atom at the Co-Gd interface compared to the Gd-Pt interface, was studied (see Table~\ref{tab:moms-cogd}, Supplementary Table 1, and Supplementary Figure 8, case C), resulting in a theoretical average spin magnetic moment of $-0.49$ $\mu_{B}$$\cdot\textnormal{atom}^{-1}$ at $T = 0$ K, which represents an enhancement of $\approx$ 3 times compared to the pristine case,  providing a possible explanation for the Gd magnetic moment values observed experimentally at low temperatures. This interpretation is supported by our experimental results in Supplementary Figure 2, which indicates a greater number of interdiffused layers at the Co-Gd interfaces compared to the Gd-Pt one. Other possible flipped spin magnetic states were also investigated; see Supplementary Note 3 for details. 
Concerning the dependence of these moments on the magnetic field, further theoretical studies are necessary to fully understand this behavior.

With further increasing $T$, the average theoretical $\bar{\mu}_{s}^{z}$ for the Pt, Co and Gd atoms at $T=300$ K, along with the experimental results obtained from XMCD, are shown in Table~\ref{tab:moms-cogd} and Figure~\ref{moms-dft}(a).
The theoretical and experimental results are in excellent agreement in both magnitude and orientation. At this temperature, the nonzero magnetization observed in the Pt/Co/Gd sample originates mainly in the Co layers, presenting a small IMM in the Pt atoms and a nonnull spin moment in Gd -- specially at the interface. Explicitly, the Pt layers follow the Co magnetization with a vanishing IMM of $\sim0.03\mu_{B}$$\cdot\textnormal{atom}^{-1}$ and the Gd layers are polarized with a large $\bar{\mu}_{s}\sim-2.74 \mu_{B}$$\cdot\textnormal{atom}^{-1}$ ferrimagnetic configuration at the Co/Gd interface, where more distant layers present small moments that tend to be antiparallel to the interfacial Gd. Because experimentally the Pt/Gd multilayer is paramagnetic at this temperature,  the Gd moment can be understood in Pt/Co/Gd as being induced as well, by proximity with Co. 
Microscopically, this influence is manifested by an antiferromagnetic exchange coupling between Co and Gd atoms, which quickly vanishes after the nearest-neighboring shell, and is responsible for maintaining a relatively large $\bar{\mu}_{s}$ at the Gd$_{4}$ even at $T=300$ K; the spin magnetic moment in the other Gd layers rapidly decreases with increasing $T$ (see Supplementary Figure 13). 
Hence, the Co/Gd interface is not only responsible for the antiparallel alignment of Gd w.r.t. Co, but also for the net magnetization in Gd observed at room temperature. These interfacial effects are instrumental to the existence of a synthetic ultrathin ferrimagnetic region that may serve as a platform for complex (non-collinear) spin structures and topologically protected states \cite{Smejkal2018,Aldarawsheh2022}.

In addition to the spin moment, another possible point of comparison with the experimental results can be made by considering the orbital magnetic moments,  $\bar{\mu}_o$. Although the Gd orbital moment cannot be directly compared with the experimental one, because it was quantified for the $4f$ states, while the theoretical treatment for Gd is acquired in the $s$, $p$ and $d$ orbitals, the Co orbital magnetic moment is a suitable candidate.  Since the orbital magnetic moment of fcc Co does not present a significant variation with temperature \cite{Menzinger1972}, a fair comparison was made between the orbital magnetic moment acquired theoretically as $\sim0.11$ $\mu_{B}$$\cdot\textnormal{atom}^{-1}$ (at $T=0$ K), and experimentally as 0.14 $\mu_{B}$$\cdot\textnormal{atom}^{-1}$ (at $T=300$ K). Therefore, the theoretical and experimental Co orbital magnetic moments are in excellent agreement.

Until now, the layer-by-layer \textit{ab-initio} analysis has shown that the magnetization is not homogeneously distributed throughout the thickness of Gd. Although the two closest layers remain in a ferrimagnetic configuration with respect to Co, the other two layers are characterized by an opposite alignment of the magnetization. To demonstrate that dynamically, in Figure~\ref{moms-dft}(b) we present ASD simulation results where the spin configuration is allowed to relax from the fully saturated state -- the equivalent of applying a sufficiently strong external magnetic field, and then suddenly removing it -- at very low temperatures ($T=10^{-3}$ K), and in the overdamped regime ($\alpha=0.5$). We see that the spin moments of Gd$_3$ and Gd$_4$ layers quickly proceed to their ferrimagnetic state with respect to Co$_1$, while, at a later stage, the spin moment $\hat{z}$-component of Gd$_2$ and Gd$_1$ reaches a minimum after $\Delta t\sim3$ ps, gradually converging to a ferromagnetic state with respect to Co$_1$. Moreover, the simulations show that the transition from the Gd saturated state to the lower-energy configuration in Pt/Co/Gd, for the range of magnetic parameters considered here, occurs dynamically in intervals of $\sim1$ ps per layer from Gd$_4\rightarrow$ Gd$_2$/Gd$_1$, characterizing an ultrafast process driven by the exchange interactions. A finite temperature analysis on Figure~\ref{moms-dft}(b), inset, shows that those intervals decrease with respect to temperature, where the transitions for each Gd layer occur in the same interval of less than 1 ps for $T=300$ K. In connection with this spin-flip dynamics, we also note that the relevant magnetization switching effect by external stimuli (e.g., femtosecond laser pulses) was mainly observed in materials involving rare-earth elements \cite{AOS-Natcomm,AOS-Nature,doi:10.1021/acsaelm.3c01534,Hintermayr2023}.

To closely inspect the microscopic origin of such a behavior of Gd, we show in Figure~\ref{moms-dft}(c) the effective exchange among Gd and the nearest Co layers, defined as $J_{\textnormal{eff}} = \sum_{i \neq j}J_{ij}$, which represents the exchange energy term in the spin Hamiltonian for a ferromagnetic (fully saturated) reference state. From this definition, $J_{\textnormal{eff}}>0$ ($J_{\textnormal{eff}}<0$) indicates the preference of parallel (antiparallel) spin alignment. 
In practice, intra- and interlayer exchange interactions were considered up to maximum cutoff radii of $\sim9.6\,$\AA/$\sim16\,$\AA, respectively. 
On the one hand, it can be seen in Figure~\ref{moms-dft}(c) that Co influence almost does not exceed the first (adjacent) layer, presenting a strong antiferromagnetic coupling for Co$_{1}$-Gd$_{4}$. 
On the other hand, Gd$_3$-Gd$_1$ and Gd$_4$-Gd$_2$ interlayer $J_{\textnormal{eff}}$ parameters are sizable and also negative, with almost the same magnitude of the neighboring Gd layers (except for Gd$_3$-Gd$_2$). This particular combination of exchange interactions, which favors longer-range Gd-Gd over Gd-Co couplings, is crucial for the preferred FSS derived from the spin Hamiltonian minimization.
 In this sense, it is worth recalling that in its bulk form (hcp), Gd exhibits a next-nearest-neighbor antiferromagnetic coupling \cite{Locht2016}, being considered as a model example of the indirect Ruderman-Kittel-Kasuya-Yosida (RKKY) interaction \cite{Scheie2022}; as is well known, this type of interaction is mediated by conduction electrons and presents a long-range nature. In contrast, fcc Co is an almost strong ferromagnet, for which a less pronouced RKKY character (as well as a faster decaying $J_{ij}$ with the $ij$ pair distance) is expected \cite{Pajda2001}.

 \begin{figure*}
\centering
\includegraphics[scale=0.8]{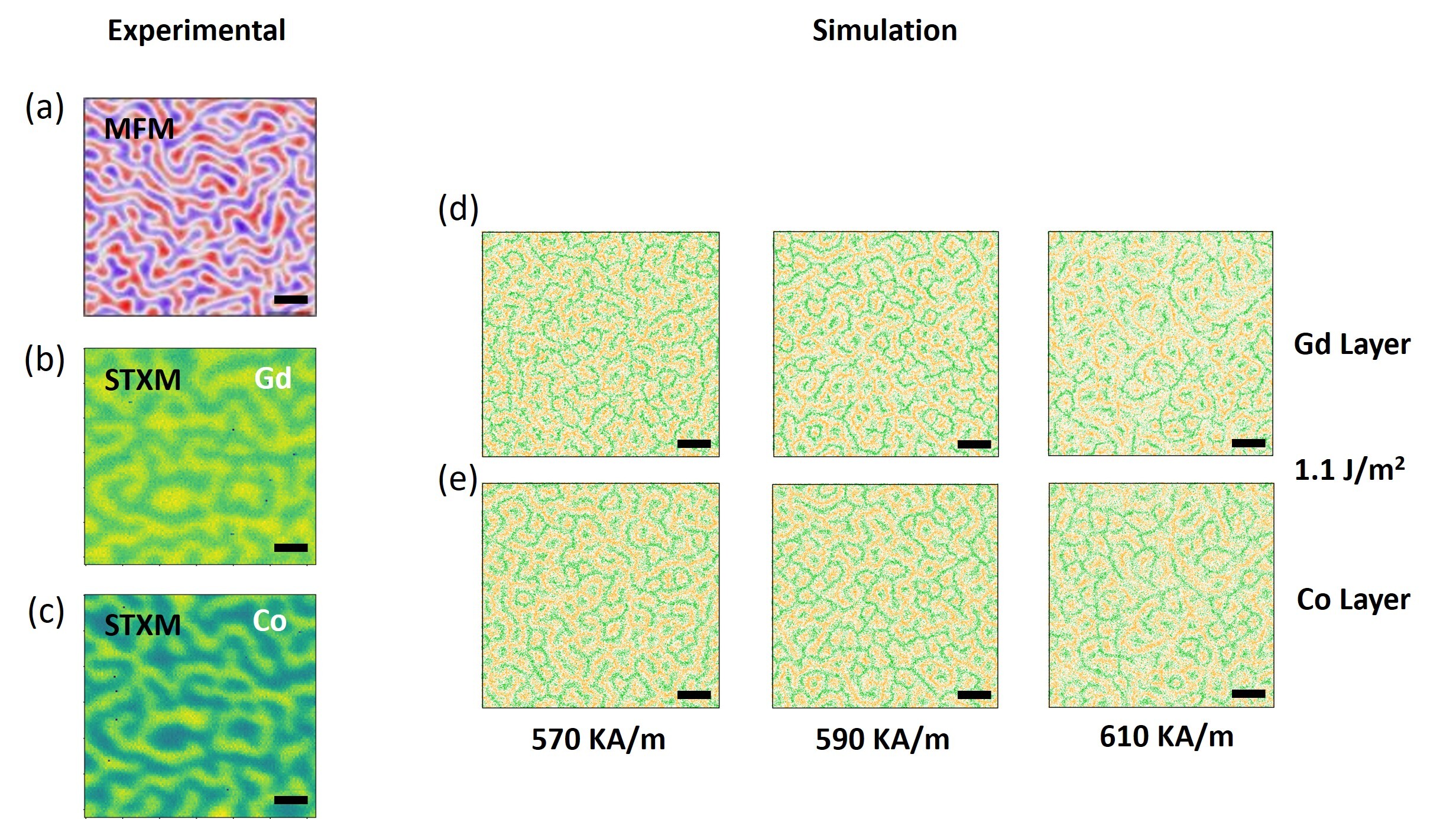}
\centering
\caption{\textbf{MFM, STXM and micromagnetic simulation images.} In (a),  magnetic force microscopy (MFM) image shows the formation of spin spirals like maze domains. In (b) and (c) scanning transmission X-ray microscopy (STXM) images performed at the L$_{3}$ Co and M$_{5}$ Gd absorption edges show the antiferromagnetic contrast of the two Co and Gd sublattices.  In (d) and (e), the micromagnetic simulation was undertaken, taking into account $D = 1.1\times10^{-3}$ J$\cdot\textnormal{m}^{-2}$ in the micromagnetic modeling. The scale bar in all images  represents 500 nm.}
\label{fig5}
\end{figure*}

ASD simulations are performed where the ($J_{ij}$, $\vec{D}_{ij}$) sets of interactions, namely the Gd$_3$-Gd$_1$ and Gd$_4$-Gd$_2$ interlayer couplings (antiferromagnetic $J_{\textnormal{eff}}$ couplings, represented in blue, in Figure~\ref{moms-dft}(c)), were artificially set to zero. In this case, all Gd layers remained with the same spin alignment. This indicates that there exists a competition between exchange interactions that energetically favors inhomogeneous magnetization throughout Gd, which is not destroyed by either the influence of thermal fluctuations at $T=300$ K (see Figure~\ref{moms-dft}(a)), or the weak magnetic anisotropy. More specifically, this behavior can be ascribed  to a combination of three connected facts: (\textit{i}) a negligible ferrimagnetic Gd-Co coupling beyond adjacent layers, linked to the (almost) strong ferromagnet character of Co; (\textit{ii}) a small $J_{\textnormal{eff}}$ coupling between neighboring Gd$_3$-Gd$_2$ layers, which can be easily perturbed by thermal fluctuations; and (\textit{iii}) the presence of long-range, RKKY-type, antiferromagnetic Gd-Gd interlayer interactions.

Finally, it is relevant to note that for the Dzyaloshinskii-Moriya interaction (DMI), although the strongest contribution comes from the intralayer $\textnormal{Co}_{\textnormal{7}}$-$\textnormal{Co}_{\textnormal{7}}$ coupling (see Supplementary Figure 11) near the Co/Pt interface, the Co/Gd interface also presents a significant interaction with the same order of magnitude, as exhibited in Figure~\ref{moms-dft}(c). The presence of DMI will be argued to play a central role in the emergence of noncollinear spin structures in the next sections.

\vspace{5mm}

\textbf{Zero-field magnetic textures at room-temperature.} Magnetic force microscopy (MFM) images were undertaken at room-temperature and zero magnetic field to observe the formation of magnetic textures in the multilayer. To directly understand the spatial magnetic configuration and correlate with the hysteresis loops, the images were performed with the sample in the remnant state after applying a magnetic field of 1 T normal to the sample surface and turning it off. The image shows spin spiral textures in the format of maze-like domains, depicted in Figure~\ref{fig5}(a).   Recalling the hysteresis loops presented in Figure~\ref{fig2}(b), the almost zero remnant magnetization agrees with the observed images, where the sample magnetization breaks into narrow domains with up and down orientation, reducing the average magnetization of the multilayer in the absence of magnetic field. 

The MFM images confirm the magnetic textures formation in the multilayer. However, the MFM cannot itself directly answer if the magnetic textures are formed in both Co and Gd layers. 
To obtain the domains' formation in each Co and Gd layer, scanning transmission X-ray microscopy (STXM) measurements were performed at the  7.0.1.2 (COSMIC) beamline of the Advanced Light Source (ALS) synchrotron. To transmit the X-ray through the sample, the same Pt(1nm)/Co(1.5nm)/Gd(1nm) multilayer was grown onto transparent silicon nitride Si$_{3}$N$_{4}$ membranes. Right- and left-circularly polarized X-rays were used, with the incoming energy fixed at the maximum of both Co L$_{3}$ and Gd M$_{5}$ absorption edge. The magnetic contrast was obtained based on XMCD, through the difference between images performed with right- and left-circularly polarization at room-temperature and zero-magnetic field.  Figure \ref{fig5}(b-c) shows the formation of spin spirals in each Gd and Co thicknesses with a very similar pattern obtained by MFM. Moreover, the contrast is opposite for each Co and Gd XMCD images, which confirms that the spin spirals texture at zero-field are coupled ferrimagnetically. These domains may be transformed in ferrimagnetic skyrmions in Pt/Co/Gd multilayer at zero-field and room temperature, by tailoring circular shaped disks with different diameters \cite{acsanm.9b01593}, or by tuning perpendicular magnetic anisotropy and remnant magnetization varying the Co thickness \cite{BRANDAO2022152598}. 

\vspace{5mm}
\textbf{Theoretical analysis: Micromagnetic simulations.}

The observation of ferrimagnetically coupled spin spirals by both MFM and STXM in our Pt/Co/Gd multilayer, with wavelengths of roughly 100 nm order of magnitude (see Fig. \ref{fig5}) makes the fully atomistic treatment unpractical. Therefore, following the philosophy of a multiscale approach \cite{Borisov2023}, the formation of those spin spirals was thoroughly investigated by, instead, performing micromagnetic simulations using Mumax$^{3}$ code \cite{Mumax} (for details, see the Methods section).
 
Figure~\ref{fig5}(d-e) summarizes the main micromagnetic results. In the bottom of Figure~\ref{fig5}(e) are labeled different values of saturation magnetization M$_{s}$ to verify how the domain structures evolve as a function of this parameter. The spin textures were stabilized in both micromagnetic Co and Gd layers. When DMI is absent in micromagnetic modeling ($D = 0$), no domain textures were stabilized, while for $D = 1.1\times10^{-3}$ J$\cdot\textnormal{m}^{-2}$  spin spirals like maze domains are obtained, as observed experimentally.  This combination of static magnetic domain images and micromagnetic simulations has been used to estimate DMI in multilayers \cite{PhysRevB.100.104430}. In addition, similar systems based on Pt/Co/HM or Pt/Co/RE (where HM is a heavy metal and RE a rare earth) multilayers have been shown to have micromagnetic DMI values around the ones used in our simulations \cite{10.1063/1.5038353,Nanoscale}.

By modifying the values of DMI, the micromagnetic simulations allowed us to qualitatively estimate this substantial interaction, where the images match the experimental findings. The DMI range of $1.1 - 1.4\times10^{-3}$ J$\cdot\textnormal{m}^{-2}$ reproduces the experimental observations very well. Furthermore, no significant differences were observed in the micromagnetic images within the M$_{s}$ interval of 570 to 610 kA$\cdot\textnormal{m}^{-1}$, as well as the J$_{\textnormal{Co-Gd}}$ interaction from $-0.1$ to $0.7\times10^{-3}$ J$\cdot\textnormal{m}^{-2}$. 
\vspace{5mm}

\textbf{Conclusions}

In conclusion, various interfacial phenomena such as induced magnetic moments, exchange coupling, and spin textures were comprehensively explored in Pt/Co/Gd multilayers by employing a multiscale approach. Our combined experimental and theoretical analysis, supported by XMCD, DFT, and ASD simulations, reveals that the Gd in proximity with Co leads to essentially two local effects: (\textit{i}) the presence of a sizable induced magnetic moment transferred from Co to Gd, which is ferrimagnetic with respect to Co at the Gd / Co interface and still survives at $T=300$ K; and (\textit{ii}) the emergence of a flipped spin state (FSS) throughout the Gd layers when the ideal fcc structure with the Pt lattice parameter is considered, and with or without interdiffusion at the interfaces. 

The FSS, characterized by a distinct alignment of Gd magnetic moments (antiparallel to Co near the interface and parallel far from it), is mainly driven by long-range (RKKY-type) Gd-Gd exchange interactions and the shorter-range influence of Co, in an environment of weak uniaxial anisotropy. The FSS across the Gd layer reduces its thickness normalized magnetic moment,   which is partly explained by the unbalanced diffusion of Gd into the Co and Pt layers and sustained by both experimental and theoretical results. It also depends on the magnetic field strength. Moreover, the Pt layers predominantly follow the magnetic moment direction of Co near the Co/Pt interface, with minimal induced moments in proximity to the Gd atoms.  

On a larger scale, the presence of the Co layers breaks the system's inversion symmetry, leading to the appearance of DMI especially at the Co/Pt and Co/Gd interfaces, via Co-Co and Co-Gd interactions, respectively. The DMI is demonstrated, via micromagnetic simulations, to play a crucial role in the formation of ferrimagnetically coupled spin spirals, observed by MFM and STXM techniques at room temperature. This observation further underscores the complexity and richness of the magnetic configurations in this system. These findings enhance our understanding
of interfacial magnetic properties in such heterostructures, placing the FSS as a mechanism to control long-distant spin transport in non-collinear antiferromagnets, as well as faster magnetization reversal in all-optical switching induced by pulsed laser.  It offers deeper insights toward the manipulation and control of antiferromagnetic coupled magnetic states, which are critical
for advancements in ultra-low-power data storage, photonics, and spintronics applications.
\vspace{5mm}

\textbf{Methods}
\vspace{5mm}

\textbf{\small {Sample fabrication}}

\small{Magnetic multilayers were fabricated by magnetron sputtering at room-temperature with a base pressure of $8\times10^{-8}$ Torr. All targets were deposited onto Si/SiO$_{2}$ substrates under Argon atmosphere pressure of $3\times10^{-3}$ Torr. First,  two reference  Pt(1nm)/Gd(1nm) and Pt(1nm)/Co(1.5nm) multilayers were grown. The bilayers were grown with 10 repetitions over a 2 nm Pt buffer layer, and to prevent oxidation, 2 nm Pt thickness was added upon the last repetition of the bilayers. Thus, the Co and Gd thicknesses were sandwiched at both sides by Pt, forming a symmetric Pt/Co/Pt and Pt/Gd/Pt heterostructure. Thus, a multilayer based on Pt(1nm)/Co(1.5nm)/Gd(1nm) with ten repeats was fabricated to study the magnetic coupling between adjacent Co and Gd atoms, as well as the formation of spin textures in both magnetic layers. The repetition number of the trilayer is justified to enhance the magnetic contrast, allowing the observation of magnetic domains in both the Co and Gd layers by scanning transmission X-ray microscopy (STXM).}

\vspace{5mm}

\textbf{\small {XAS and XMCD}}

X-ray absorption (XAS) and X-ray magnetic circular dichroism (XMCD) were performed around the Co, Gd, and Pt absorption edges at the Brazilian Synchrotron Light Laboratory (LNLS). XAS and XMCD undertaken for Co and Gd were measured in the Soft X-ray Absorption and Imaging (SABIÁ) beamline located in the SIRIUS, using a superconducting magnet which allows applying magnetic fields up to 9 T perpendicular to the sample surface. The system is also coupled with a variable temperature insert (VTI) where the sample is mounted and the temperature varied. The signal was recorded by total electron yield mode in the fast flyscan acquisition. In order to exclude variations from the X-ray intensity coming from the delta undulator, a reference signal was measured by the transmission of the X-ray  through the gold grid before the beam illuminates the sample. Thus, the XAS spectra acquired with right- and left-circular polarization were normalized by the reference.  

XAS and XMCD were acquired for Pt in the Extreme Conditions and Analyses (EMA) beamline, also located in the Sirius. The measurements were acquired at room temperature and applied a magnetic field of 1 T normal to the sample surface. Using a one-quarter plane wave, the X-ray polarization was modified from right- and left-circular polarization. The measurements were acquired in total fluorescence yield (TFY) using a step scan.

\textbf{{Theoretical multiscale approach:}}\\

\textbf{{1: Atomistic spin dynamics}}\\

\small{To perform first-principles calculations the
the real-space linear-muffin-tin-orbital within the atomic sphere approximation (RS-LMTO-ASA) method \cite{Pessoa1992}, based on density functional theory (DFT) \cite{HohenbergKohn1964} was used. The ground state electronic density was obtained by using the Haydock recursion \cite{Haydock1980}, with the recursion cut-off of $LL=22$, in addition to the Beer-Perttifor terminator \cite{Beer1984} and the local spin density approximation (LSDA) \cite{Barth1972} as the exchange-correlation functional.
The atomistic spin simulations were done using the Uppsala Atomistic Spin Dynamics (UppASD) \cite{Eriksson2017} code, where the Landau-Lifshitz-Gilbert (LLG) equation \cite{Gilbert1955} is solved to obtain the magnetic configuration with minimum energy.}

An infinite geometry with perfect \textit{fcc} stacking in the [111] direction, composed of 4 layers of Pt, 4 layers of Gd and 7 layers of Co (see Figure~\ref{moms-dft}(a)) was considered.
The DFT calculations were performed using the Real Space Linear Muffin-Tin Orbital Atomic Sphere Approximation (RS-LMTO-ASA) method \cite{Pessoa1992,Peduto1991,Klautau1999,Rodrigues2016,BezerraNeto2013,Cardias2016,Kvashnin2016,Cardias2020,PhysRevB.70.193407,KLAUTAU200527,PhysRevB.66.132416,KLAUTAU2002385} to obtain the magnetic parameters, including magnetic moments and exchange constants, namely the isotropic exchange coupling ($J_{ij}$) and the Dzyaloshinskii-Moriya vectors ($\vec{D}_{ij}$). A deeper analysis on the structure and the ground state search using different DFT methods can be found in the Supplementary Note 3. Subsequently, the obtained \textit{ab-initio} values were used to parameterize the spin Hamiltonian and solve the phenomenological Landau-Lifshitz-Gilbert (LLG) equation of motion, with temperature variations from $\sim0$ to $300$ K, via the Uppsala Atomistic Spin Dynamics (UppASD) code \cite{Antropov1996,Eriksson2017,Skubic2008}.

Concerning the ASD simulations, the inclusion of induced magnetic moments can be crucial for the determination of $T_{\textnormal{C}}$ and the magnetic properties at finite temperatures of FM/heavy metal systems \cite{PhysRevB.82.214409}. However, since the IMM for Pt layers is only sizable 
at the interface Pt/Co (as discussed below, see Figure~\ref{moms-dft}(a)), including further Pt atoms with vanishing small magnetic moments in the classical ASD approach is not appropriate \cite{JMathPhys.12.1000}. Then, here, we considered a hexagonal lattice of $200\times200\times n$ spins, where \textit{n} is the number of layers, considering one Pt monolayer at the interface Pt/Co followed by all the Co and Gd layers, resulting in $n=12$ (see Supplementary Note 3 for information regarding the range of interactions considered).
The temperature is increased continuously from $0$ K to $300$ K, where the atomic moments present a deviation of the $\hat{z}$-axis due to thermal fluctuations. As the XMCD results consider contributions from all the layers of a given chemical species (for Co and Gd, only the repetitions closest to the sample surface due the TEY mode detection, see Figure~\ref{fig2}(a)) \cite{STOHR1995253}, a fair comparison to the theoretical values can be given by the average projection of the atomic spins to the spin quantization axis ($\hat{z}$-axis), $\bar{\mu}_s^{z}$.  
For the cases in which a finite temperature ($T>0$) is involved, these values are obtained by considering the dynamical average $\hat{z}$-component of $\vec{\mu}_{s_i}$ for all sites $i$ in a given layer, for a sufficient ASD simulation time of $t=10$ ps.
\\

\textbf{2: Micromagnetic simulations: Modeling }

Within the micromagnetic approach, to model our system, using the Mumax$^{3}$ code \cite{Mumax}, two layers representing Co and Gd were geometrically divided into cell sizes of 0.3 nm along the  $\hat{x}-\hat{y}$ plane, and 0.5 nm out-of-plane ($\hat{z}$). The total area along $\hat{x}-\hat{y}$ is 2500$\times$2500  nm$^{2}$, and Co and Gd thicknesses 1.5 and 1 nm along the $\hat{z}$ direction, respectively. The simulations were carried out at $T=300$ K, zero field and during $t=10$ ns.  

The exchange stiffness constant was fixed to J$_{\textnormal{Co-Co}}$ = 1.5$\times$10$^{-12}$ J$\cdot\textnormal{m}^{-3}$ and J$_{\textnormal{Gd-Gd}}$ = 0.5$\times$10$^{-12}$ J$\cdot\textnormal{m}^{-3}$~\cite{natcomm}. 
The antiparallel alignment between the Co and Gd layers in the interface was ensured by using interlayer exchange coupling (IEC),  mimicking an interfacial exchange J$_{\textnormal{Co-Gd}}$ interaction for the nearest Co and Gd atoms. The J$_{\textnormal{Co-Gd}}$ was initially fixed in $-0.5\times$10$^{-3}$ J$\cdot\textnormal{m}^{-2}$.  The PMA obtained experimentally was fixed in  0.12 MJ$\cdot\textnormal{m}^{-3}$  and saturation magnetization modified from 570 to 610 kA$\cdot\textnormal{m}^{-1}$ to account for any variation of this parameter in formation of the magnetic texture. As the atomistic results previously discussed suggest, a sizable DMI is also present due to the broken inversion symmetry structure. Thus, a DMI constant $D$ was also considered in the simulations and varied together with the interfacial exchange J$_{\textnormal{Co-Gd}}$ to explore their impact on the spin textures emergence. Note here we differentiate the atomistic parameters $J_{ij}$ and $\vec{D}_{ij}$ used in the ASD simulations, from the effective parameters $J$ and $D$ used in the micromagnetic simulations.

Additional results of micromagnetic simulations are described in the Supplementary Note 2. 

\vspace{5mm}

\begin{acknowledgement}

J.B, J.C.C, F.B, T.J.A.M, H.M.P, A.B.K, and P.C.C. acknowledge financial support from   Brazilian agencies CAPES, CNPq, FAPESP and FAPESPA. In particular, São Paulo Research Foundation (FAPESP) is acknowledged by financial support through the processes: \#2020/05609-7 and \#2022/10095-8 (P.C.C., A.B.K. and H.M.P),   \#2012/51198-2 (J.C.C). 
A.B.K. acknowledges support from 
the INCT of Materials Informatics, and A.B.K., J.B., T.J.A.M., F.B.  and J.C.C. acknowledge the INCT of Spintronics and Advanced Magnetic Nanostructures (INCT-SpinNanoMag).  
A.B. acknowledges eSSENCE. A.B. and I.P.M. acknowledge support from the Knut and Alice Wallenberg Foundation. F.B acknowledges support from the CNPq project with process number 421070/2023-4.
XAS and XMCD measurements were carried out around the Co and Gd absorption edges in the soft X-ray absorption and imaging (SABIÁ) beamline of the Brazilian Synchrotron Light Laboratory (LNLS/CNPEM), as well as the extreme condition analyzes (EMA) beamline. Previous XAS and XMCD were also performed in the former UVX, which was also localized in the LNLS/CNPEM. The calculations were performed at the computational
facilities of the National Laboratory for Scientific Computing (LNCC/MCTI, Brazil), CCAD-UFPA (Brazil), and at the National Academic Infrastructure for Supercomputing in Sweden (NAISS), partially funded by the Swedish Research Council through grant agreement no. 2022-06725.

\end{acknowledgement}

\vspace{5mm}


\textbf{Supporting information}

\small{Additional description of the experimental and theoretical details, including: (\textit{i}) X-ray diffraction to identify the growth direction of the thin films, X-ray absorption and the sum rules used to obtain the orbital and spin magnetic moments, and (\textit{ii}) the computational procedure to perform first principles calculations based on density functional theory (DFT), and atomistic spin dynamics. In addition, further information on the micromagnetic simulations is presented. }

\vspace{5mm}

\textbf{Competing interests} 

All authors declare no competing interests.
\vspace{5mm}

\textbf{Author contributions} 

J.B. conceived the experimental idea of the projetc. I.P.M., A.B.K., and H.M.P. designed the theoretical project.  P.C.C, I.P.M., J.B., and A.B.K. carried out the simulations and calculations. I.P.M. and A.B. developed new implementations on RS-LMTO-ASA code necessary to develop this project. J.B., T.J.A.M., and J.C.C. performed the XAS and  XMCD experiments. T.J.A.M. undertook STXM images. F.B. and J.B measured hysteresis loops. The initial version of the manuscript was written by J.B., P.C.C., I.P.M., and A.B.K. A.B., H.M.P., and J.C.C. discussed the results and commented on the manuscript. All authors contributed to discussions, writing, and revision of the manuscript to its final version.

\vspace{5mm}

\textbf{Data availability} 

Most data needed to reproduce the results are available in the Supplementary Notes. Additional details supporting the findings of this study can be provided upon reasonable request from the corresponding authors (J.B. and I.P.M.).

\vspace{5mm}

\textbf{Code availability} 

The codes used in the theoretical part (RS-LMTO-ASA, UppASD, and Mumax$^3$) are described and referenced in the Methods section and are available free-of-charge.

\bibliography{Ref-Final.bib}

\end{document}